\documentclass[runningheads]{llncs}

\usepackage{colortbl}
\definecolor{mygreen}{RGB}{144,238,144}

\usepackage{tikz}
\usepackage{tikzscale}
\usepackage{pgfplots}
\usetikzlibrary{pgfplots.groupplots}
\usepackage[hidelinks]{hyperref}

\newlength\figureheight
\newlength\figurewidth

\usepackage[T1]{fontenc}

\usepackage{graphicx}

\usepackage[title]{appendix}

\usepackage{longtable}

\usepackage{amsmath}

\usepackage[bibstyle=ieee, maxcitenames=2, mincitenames=1, maxbibnames=999999, sorting=none, natbib=true, doi=false,isbn=false,eprint=false]{biblatex}
\usepackage{csquotes}
\pgfplotsset{compat=1.18} 
\begin{document}

\raggedbottom
\title{An End-to-End Workflow using Topic Segmentation and Text Summarisation Methods for Improved Podcast Comprehension}

\author{Andrew Aquilina\inst{1}
\and
Sean Diacono\inst{1}
\and
Panagiotis Papapetrou\inst{1}
\and
Maria Movin\inst{1,2}
}

\titlerunning{Topic Segmentation and Text Summarisation for Podcast Data}
\authorrunning{A. Aquilina, S. Diacono, \textit{et al.}}

\institute{Department of Computer and Systems Sciences, Stockholm University, Sweden\\
\email{\{anaq3720,sedi5808\}@student.su.se}, 
\email{\{panagiotis,maria.movin\}@dsv.su.se}
\and Spotify AB, Sweden\\
%\textsuperscript{a}\email{\{anaq3720,sedi5808\}@student.su.se} \\
%\textsuperscript{b}\email{\{panagiotis,maria.movin\}@dsv.su.se}
}
\maketitle              % typeset the header of the contribution

\begin{abstract} 
The consumption of podcast media has been increasing rapidly. Due to the lengthy nature of podcast episodes, users often carefully select which ones to listen to. Although episode descriptions aid users by providing a summary of the entire podcast, they do not provide a topic-by-topic breakdown. This study explores the combined application of topic segmentation and text summarisation methods to investigate how podcast episode comprehension can be improved. We have sampled 10 episodes from Spotify's English-Language Podcast Dataset and employed TextTiling and TextSplit to segment them. Moreover, three text summarisation models, namely T5, BART, and Pegasus, were applied to provide a very short title for each segment. The segmentation part was evaluated using our annotated sample with the $P_k$ and WindowDiff ($WD$) metrics. A survey was also rolled out ($N=25$) to assess the quality of the generated summaries. The TextSplit algorithm achieved the lowest mean for both evaluation metrics ($\bar{P_k}=0.41$ and $\bar{WD}=0.41$), while the T5 model produced the best summaries, achieving a relevancy score only $8\%$ less to the one achieved by the human-written titles.

\end{abstract}

\section{Introduction}

The consumption of podcast media has been increasing rapidly. In 2020, an estimate of 155 million users were listening to podcasts every week, while the total of monthly podcast US consumers has grown by 16\% year-over-year \cite{edison}. This has encouraged media service providers, such as Spotify, to reconsider how users could be provided with such content. An example of how this could be achieved is by providing a deeper context of the ‘topics’ discussed in a podcast episode. In podcasts, a topic generally refers to what is being conveyed in the discourse. More often than not, podcast episodes shift from one topic to another, each revolving around keywords of interest. These are essential in understanding the flow of information, but they are simply not enough for listeners to grasp further context. For example, assume a piece of dialogue explaining the healthy benefits of fruit. The keyword ‘fruit’ explains less than the summary of ‘why fruits are good for you’. Therefore, summarising segments into short titles revolving around specific topics opens the door for improved and concise comprehension.

One of the challenges pertaining podcasts is their lengthy nature. This incentivises users to carefully select which episode to commit to before listening, ensuring that the provided content would be relevant or interesting to them. In fact, surveys reveal that listeners pay considerable attention to the text description of a podcast before deciding whether to listen to it\footnote{\url{https://www.thepodcasthost.com/promotion/podcast-discoverability/}}. Unfortunately, the utility of descriptions falls short if users are looking to consume specific parts of an episode. For example, if one wants to listen to a particular story or discussion, the listener is unable to do so without going over the entire episode. However, manually dividing transcripts may require skimming the full dialog, making it a very laborious and time-consuming task. It may also prove itself as a non-obvious task for human annotators. A widely used method to address such a problem is topic segmentation, which aims to automatically split a single document into shorter, topically coherent segments by an objective algorithm.

To combat such challenge, the newly proposed TREC 2020 Podcast Track \cite{jones2021trec} encouraged research into two isolated tasks, namely segment retrieval and summarisation. Text summarisation systems focus on generating a short excerpt describing the contents of an entire podcast episode. While these excerpts provide the user with podcasts pertinent to a given search query, they do not deliver a deeper context by labelling the topically-distinctive parts of the episode. As a result, information revealing the beginning or end to a topic is lost. Overcoming this can be achieved through segmented summaries, enabling listeners to not only navigate podcasts with ease, but also improve comprehension, searching capabilities, and information access. All in all, the lack of comparative studies for the combination of both tasks becomes evident. By studying the effects of summarising topically segmented podcasts into short titles and reporting whether this improves episode comprehension contributes to filling the outlined knowledge gap. Providing potential solutions could also shed light on how information access and users' comprehension of podcast content can be improved. 

The main aim of this paper is to therefore investigate the combined application of topic segmentation and text summarisation methods to improve podcast episode comprehension, sampling podcast episodes from Spotify's English-Language Podcast Dataset \cite{dataset}. The structure of podcast data presents unique challenges in the pursuit of such an aim. For example, representing podcast as text can lead to potential inaccuracies caused by automatic speech recognition (ASR) methods. The structure of spoken and written language also vary significantly \cite{power2003document}. We therefore propose a workflow that includes the following contributions: (i) We evaluate the predictive performance of topic segmentation methods when compared to manually annotated segments, (ii) We carry out a preliminary survey using a small sample of episodes and users to determine whether titles generated by text summarisation methods encapsulate the topics of the identified segments, especially when compared to human-written titles. 

\newpage

\section{Background \& Related Work}

\subsection{Topic Segmentation}

The aim of topic segmentation is to divide blobs of text into semantically coherent segments, either to enhance human interpretation or to facilitate Natural Language Processing applications. As noted from the survey of \citet{purver2011topic}, related work on topic segmentation can be grouped into two categories: either through the specific detection of topic transitions (from the use of cue words or other prosodic features), or by noting changes in the text’s vocabulary. Given the scope of this study, we shall focus on the latter, namely on discriminative and clustering-based methods.

Topic Segmentation efforts were founded on the intuition that topical shifts are characterised by vocabulary changes. Therefore, by detecting alterations in lexical use, topic boundaries can be detected. Making use of such an intuition is the TextTiling algorithm \cite{hearst1997text}, one of the most well-known methods within such a domain. The algorithm is prominently used and built upon to this day. An example of this would be TopicTiling \cite{riedl2012topictiling}, which employs Latent Dirichlet Allocation (LDA) and assigns a topic to each word to aggregate topic-count for fixed-size windows. \citet{he2020improvement} also improve the TextTiling algorithm by introducing a curve-smoothing process, further highlighting the topic changes within segments. The algorithm has also been enhanced through the use of semantic word embeddings, improving on benchmarked approaches \cite{gupta2020comparative} .

Instead of detecting points of low cohesion as to note topic changes, clustering-based methods group highly cohesive sentences to reduce the text into a number of topics. The interest in unsupervised techniques is prominenet for clustering-based topic segmentation methods, and the work of \citet{alemi2015text} is no exception, giving birth to Content Vector Segmentation (CVS). By employing GloVe word embeddings, CVS iteratively segments text through the generation of segment scores, splitting them into smaller groups until a threshold is reached. The authors demonstrate state-of-the-art performance. The TextSplit algorithm \cite{textsplit} was recently developed based on the aforementioned work. \citet{nangi2019offvid} employed TextSplit to produce topic-specific video segments to index video lectures. Their work was evaluated through a questionnaire-based survey.

According to \citet{jing2021identifying}, there has been a scarcity of topic segmentation efforts specifically on spoken-word content and podcast data. Their contemporary work focused on identifying the introductory section from a manually-annotated dataset of 400 podcast episodes. The authors trained three Transformer models based on the pre-trained BERT model, providing a basis for introduction segmentation on podcast data. Earlier efforts by \citet{fuller2008using} aimed to enable users skimming podcast episodes. The authors found that the TextTiling algorithm agreed quite well with the human-made segments. Their work was conducted on a corpus of 30 podcast episodes.

\subsection{Text Summarisation}

Text Summarisation is the process of taking a piece of text, selecting the most important information, and creating an abridged version. As manual text summarisation is arduous and time consuming, automatic summarisation systems have been developed. Such approaches can be classified into two main groups: extractive methods, where important sentences are directly taken from the source text and joined together to form a summary, and abstractive methods, where new sentences are generated to summarise the salient parts. For the scope of this study, we have focused on abstractive summarisation systems using deep-learning based methods.

Deep-learning based methods use Recurrent Neural Network (RNN) Sequence-to-Sequence (Seq2Seq) models to generate summaries \cite{gupta2019abstractive}. One such architecture is BART \cite{lewis2019bart}, achieving state-of-the-art results in text generation tasks. \citet{lewis2019bart} fine-tuned BART on summarisation datasets and achieved Recall-Oriented Understudy for Gisting Evaluation 1 (ROGUE-1) scores of $44.16$ on the CNN/DailyMail (CNN/DM) dataset \cite{nallapati2016abstractive}.  Alternatively, Pegasus \cite{zhang2020pegasus} is a model with a pre-training task aimed at abstractive text summarisation. Fine-tuned for text summarisation, Pegasus attained a ROGUE-1 score of $37.68$ on the Annotated Enron Subject Line Corpus (AESLC) dataset \cite{zhang2019email}. Another deep-learning model used for text summarisation is the Text-To-Text Transfer Transformer (T5) model \cite{raffel2020exploring}. Fine-tuned on the CNN/DM dataset, the T5 architecture achieved a ROGUE-1 score of $43.52$. 

Several techniques have been applied to podcast data to explore their effectiveness. \citet{rezapour2020spotify} used BART for abstractive summarisation and achieved an F-measure score of $18.42\%$ compared to human-written descriptions. The models were pre-trained on the CNN/DM
dataset and then fine-tuned on Spotify’s English-Language Podcast Dataset. Their system outperformed humans according to a survey. \citet{karlbom2020abstractive} combined BART with Longformer Attention and achieved a higher F-measure score of $19.23\%$. Although these abstractive summarisation systems generate coherent summaries, they are usually still made up of multiple sentences, making them too lengthy to address our research problem.

Research has been done to apply text summarisation techniques for headline and title generation. Such approaches are able to create summaries made up of just a few words for a piece of text, making them ideal for tasks such as news headline generation and scientific paper title generation. The Gigaword dataset has enabled the development of such short summarisation systems by being made up of news articles and their respective headlines \cite{graff2003english}. \citet{aghajanyan2021muppet} proposed a language generation model based on BART. Their research focused on fine-tuning the model using different language generation datasets. The model achieved a ROGUE-1 score of $40.40$ on the Gigaword dataset, surpassing the previous Pegasus model's score of $39.12$. These models, although untested on podcast data, are promising candidates for our study's second objective.

\subsection{Segmentation-based Summarisation}

There has been prior work combining topic segmentation and text summarisation efforts. \citet{cho2022toward} investigated the utility of segmentation methods for extractive summarisation of lengthy scientific articles. Their approach learns representations using a Longformer model to conduct segmentation and summarisation simultaneously.  Similarly, \citet{liu2021end} build two text segmentation models to also improve the extractive summarisation task, noting that a segmentation-based approach enhances summarisation quality especially when information is not found at the beginning of a document. Their text segmentation model is based on a modified version of TextTiling using BERT embeddings. Experimenting with both extractive and abstractive summarisation systems, \citet{miculicich2023document} propose two frameworks to summarise news articles. The first is the \textit{Pipeline} framework, which first segments the text using a segmentation model, and then applies a title generation model to generate the headings. The other is the \textit{Joint} approach, which tackles both tasks using a single encoder-decoder neural model. The authors note the beneficial nature of the latter as it allows the previously generated headings to be known when generating the concurrent title. While this may be advantageous for news articles due to their high level of topical cohesion, this may not be the case for podcast data.

\section{Methodology}

\subsection{Problem Formulation}
Let $\mathcal{P}=\{P_1, \ldots, P_n\}$ define a set of $n$ podcasts, where each podcast is described by a transcribed piece of text, which we refer to as the \emph{podcast transcript}. Moreover, for each podcast $P_i\in \mathcal{P}$, we can find a set of $k_i$ indices $\mathcal{J}_i=\{\beta_1, \ldots, \beta_{k_i}\}$ that in turn defines a set of $k_i$ segments $f = \{P_i[1:\beta_1], P_i[\beta_1+1:\beta_2], \ldots, P_i[\beta_{k_i-1}+1:\beta_{k_i}]\}$, with $\beta_{k_i}=|P_i|$, i.e., the last segment ends at the end of each podcast. Each podcast segment can also be described by some text summary, which encapsulates the main discussion point of said segment. In addition, for each podcast $P_i\in \mathcal{P}$, there exists $\mathcal{F}_i = \{f_1, \ldots, f_{m_i}\}$, which defines a set of $m$ possible topic segmentations, and $\mathcal{G}_i = \{g_1, ..., g_{l_i}\}$ which defines a set of $l$ possible text summaries given these segments. 

For each podcast $P_i \in \mathcal{P}$, we want to find a set of segments $f^*$ from $\mathcal{F}_i$ and a set of summaries $g^*$ from $\mathcal{G}_i$, such that $f^*$ yields the minimum error for topic segmentation and $g^*$ yields the maximum score for average summary relevancy. Let $\mathcal{E}_1$ and $\mathcal{E}_2$ be defined as evaluation functions for $f$ and $g$ respectively\footnote{Instantiations for both functions can be found and further explained in Sections 4.2 and 4.3.}. The problem is denoted in Equation 1.

\begin{equation}
    \forall P_i \in \mathcal{P} \text{, such that } f^* = \operatorname*{argmin}_{f\in\mathcal{F}_i} \mathcal{E}_1(f) \text{ and } g^* = \operatorname*{argmax}_{g\in\mathcal{G}_i} \mathcal{E}_2(g) 
\end{equation}

\subsection{Workflow overview}

\begin{figure}[!htb]
		\centering
		\includegraphics[width=\linewidth]{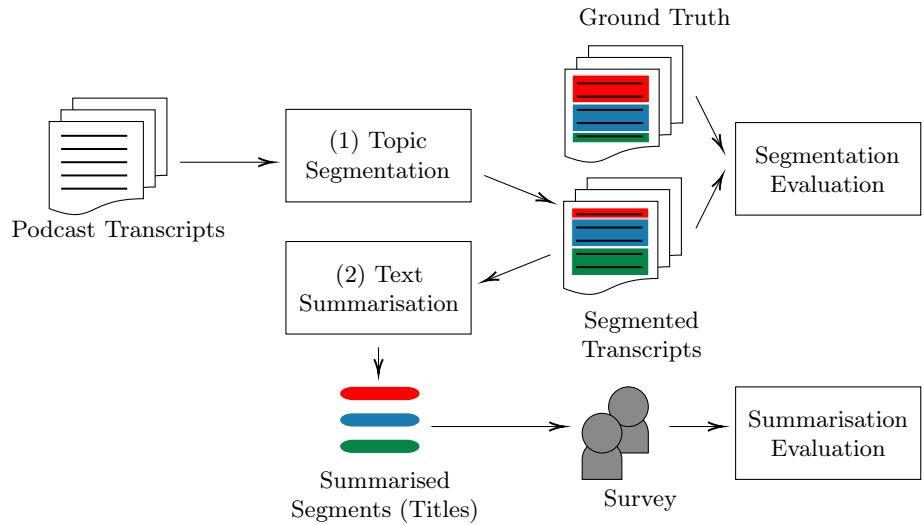}
		\caption{An overview of our proposed workflow.}
		\label{fig:workflow}
\end{figure}

The main steps of the proposed workflow addressing our research problem are depicted in Figure \ref{fig:workflow}. We have taken cues from the \textit{Pipeline} approach, as outlined by \citet{miculicich2023document}, to develop our methodology. Our workflow mostly consists of two steps: (1) topic segmentation, and (2) text summarisation. Sampled podcast transcripts are segmented using the topic segmentation algorithms, namely TextTiling and TextSplit, followed by an evaluation using human-made segments to report how well the podcasts' topics are captured. The segments stemming from the best performing topic segmenter are then summarised using the text summarisation models, namely T5, BART, and Pegasus, to provide a title to each segment. A survey is then deployed to a number of human participants to evaluate the relevancy of the generated summaries. To our knowledge, we are the ﬁrst to utilise and provide a comparative evaluation of the aforementioned topic segmentation algorithms in conjuction with such text summarisation methods to improve podcast episode comprehension.

\subsection{Topic Segmentation}

Building on the majority of prior research \cite{song2016dialogue, xu2021topic}, we utilised the implementation of the TextTiling algorithm as provided by the \texttt{NLTK} module. The TextTiling algorithm \cite{hearst1997text} divides the given text into equally sized windows and represents each with a lexical frequency vector. Cosine similarities of adjacent windows are then calculated to hypothesise topic boundaries. The algorithm’s parameters consist of the pseudosentence length $w$, block size $k$, as well as the chosen approach to threshold the depth score plot $f$. While the author of the algorithm suggests optimal values for such parameters \cite{hearst1997text}, namely $w$ = 20 and $k$ = 10, we conduct a grid-search to tune the algorithm's parameters. Our intention is to explore the parameter space, tuning the discussed parameters within the context of the podcast transcripts. 

We also compare the TextTiling algorithm with TextSplit, an implementation based on the clustering-based segmentation technique of \citet{alemi2015text}. By representing sentences in vector space through the employment of word embeddings, sentences are clustered together to form a number of topics. The algorithm terminates when the number of topics reaches the predefined penalty threshold $p$ or when the reached gain from performing a split is below the required amount. There exist two variants of this algorithm, greedy or dynamic. The penalty hyper-parameter $p$ defines the granularity of the split. While the greedy approach iteratively splits the text until it is below the given penalty threshold, the dynamic alternative finds the optimal $p$ value given a maximum number of sentences that belong to a segment. We conduct a tuning experiment to note which segment-length $l$ value yields the best $p$ value. 

\subsection{Text Summarisation}

We made use of several text summarisation models to give short titles to each segment extracted from the podcast transcripts. Through a Python package, the HuggingFace Transformer's open-source library makes available several summarisation models, such as T5, BART, and Pegasus. This open-source library allows for inference with models fine-tuned on text summarisation datasets, hence why it was used in our research. The T5 architecture was implemented due to its flexibility and potential for good performance on our text summarisation task. The T5 model chosen from the HuggingFace library was fine-tuned on a WikiNews dataset made up of 500 thousand articles and their headlines. Based on preliminary qualitative testing using the HuggingFace interface, this model was found to produce appropriate titles for podcast segments. For this reason, we included this model in our research for further evaluation. The second model incorporated in this study is a Pegasus model fine-tuned on the AESLC dataset. The AESLC dataset is made up of email bodies and their respective subject lines \cite{zhang2019email}, making it ideal for the text summarisation task. As discussed in Section 2, the Pegasus architecture achieved state-of-the-art results on this dataset \cite{zhang2020pegasus} and it is included in the HuggingFace library, hence why it has been included. The BART architecture is the final text summarisation technique investigated in this research. The BART model used from the HuggingFace library was fine-tuned on the X-Sum dataset \cite{narayan2018dont}. This dataset is comprised of news articles and summaries made up of just one sentence. Therefore, this version of BART generates extracts that also fit the requirements of our text summarisation task. We include Table \ref{app:tab:gen_sum} in the Appendix to present a collection of generated titles as examples.

\section{Evaluation}

\subsection{Dataset}
The dataset used in this study is the English-Language Podcast Dataset from Spotify \cite{dataset} that includes both audio and text data. The corpus comprises of 100,000 podcast episodes varying in production quality, topics, and structural formats. For each episode, a transcript captured through the means of ASR is provided.  For the scope of this study, we have randomly sampled a subset of 10 episodes and focused solely on such transcripts. Metadata, such as the show's name and publisher's area, are also included for each episode, but topical segments are not. As we require such segmentations to answer our research question, the sampled dataset has been annotated with segment boundaries. This was done by listening to the episodes themselves and marking down topic shifts.

\subsection{Topic Segmentation}

\subsubsection{Adopted Evaluation}

% As described by \citet{purver2011topic}, the $P_k$ and $WD$ metrics were used to evaluate the text segmentation algorithms’ performance when compared to the human reference points. We also evaluate their performance to a random baseline, starting a new segment with probability $\frac{1}{k}$, where $k$ is the number of average segments in the annotated set. For this, we used the \texttt{SegEval} Python package\footnote{\url{https://bit.ly/3M8bFIv}}. The best segmentation technique according to these metrics was utilised within the second objective.

As described by \citet{purver2011topic}, the $P_k$ and $WD$ metrics were used to evaluate the algorithms’ performance when compared to the human reference points. $P_k$ is calculated as follows. Let $\delta_S(i, j)$ indicate whether sentences $i$ and $j$ are in distinct segments, evaluating to $1$ if so and $0$ in the opposite case. The type $S$ refers to whether the segmentation is derived from reference $R$ or hypothesis $H$. Therefore $\delta_H(i, j) \oplus \delta_R(i, j)$ denotes whether $H$ and $R$ disagree about the separation of $i$ and $j$, where $\oplus$ is the XOR operator. $P_k$ is obtained by moving a sliding window across the entire segments, aggregating this score, and dividing by the number of windows $N-k$, shown in Equation \ref{eqn:pk}.

\begin{equation}
    P_k = \frac{\sum^{N-k}_{i=1}\delta_H(i, i+k) \oplus \delta_R(i, i+k)}{(N-k)}
    \label{eqn:pk}
\end{equation}

The $WD$ metric is calculated in a similar fashion, shown in Equation \ref{eqn:wd}. While a fixed-width window is also moved across the data, windows are classified as "correct" if and only if the same number of boundaries are assigned between their start and end. This is defined with the variable $b_S(i, j)$.

\begin{equation}
    WD = \frac{\sum^{N-k}_{i=1}[|b_H(i, i+k) - b_R(i, i+k)|>0]}{(N-k)}
    \label{eqn:wd}
\end{equation}

We also evaluate their performance to a random baseline, starting a new segment with probability $\frac{1}{k}$, where $k$ is the number of average segments in the annotated set. For this, we used the \texttt{SegEval} Python package \cite{Hiroki_seg_eval}. The best segmentation technique according to these metrics was utilised within the second objective.

\subsubsection{Results}

\begin{figure}[!htb]
		\centering
		\setlength\figureheight{4.6cm}
		\setlength\figurewidth{0.5\linewidth}
		% This file was created with tikzplotlib v0.10.1.
\begin{tikzpicture}

\definecolor{darkgray176}{RGB}{176,176,176}
\definecolor{darkorange25512714}{RGB}{255,127,14}
\definecolor{lightblue}{RGB}{173,216,230}
\definecolor{lightgreen}{RGB}{144,238,144}
\definecolor{pink}{RGB}{255,192,203}

\begin{groupplot}[group style={group size=2 by 1}]
\nextgroupplot[
height=\figureheight,
width=\figurewidth,
tick align=outside,
tick pos=left,
title={$P_k$},
x grid style={darkgray176},
xmin=0.5, xmax=3.5,
xtick={1, 2, 3},
xticklabels={Baseline, TextTiling, TextSplit},
y grid style={darkgray176},
ylabel={Scores},
ymajorgrids,
ymin=0.239109181552262, ymax=0.605784609372493,
ytick style={color=black}
]
\path [draw=black, fill=pink]
(axis cs:0.85,0.411811112189385)
--(axis cs:1.15,0.411811112189385)
--(axis cs:1.15,0.522353104561343)
--(axis cs:0.85,0.522353104561343)
--(axis cs:0.85,0.411811112189385)
--cycle;
\addplot [black]
table {%
1 0.411811112189385
1 0.313609467455621
};
\addplot [black]
table {%
1 0.522353104561343
1 0.589117544471573
};
\addplot [black]
table {%
0.925 0.313609467455621
1.075 0.313609467455621
};
\addplot [black]
table {%
0.925 0.589117544471573
1.075 0.589117544471573
};
\path [draw=black, fill=lightblue]
(axis cs:1.85,0.425947396029756)
--(axis cs:2.15,0.425947396029756)
--(axis cs:2.15,0.480744634829876)
--(axis cs:1.85,0.480744634829876)
--(axis cs:1.85,0.425947396029756)
--cycle;
\addplot [black]
table {%
2 0.425947396029756
2 0.405594405594406
};
\addplot [black]
table {%
2 0.480744634829876
2 0.530673854447439
};
\addplot [black]
table {%
1.925 0.405594405594406
2.075 0.405594405594406
};
\addplot [black]
table {%
1.925 0.530673854447439
2.075 0.530673854447439
};
\addplot [black, mark=o, mark size=3, mark options={solid,fill opacity=0}, only marks]
table {%
2 0.255776246453182
};
\path [draw=black, fill=lightgreen]
(axis cs:2.85,0.395316507996345)
--(axis cs:3.15,0.395316507996345)
--(axis cs:3.15,0.429922973634411)
--(axis cs:2.85,0.429922973634411)
--(axis cs:2.85,0.395316507996345)
--cycle;
\addplot [black]
table {%
3 0.395316507996345
3 0.362402303578774
};
\addplot [black]
table {%
3 0.429922973634411
3 0.4631507775524
};
\addplot [black]
table {%
2.925 0.362402303578774
3.075 0.362402303578774
};
\addplot [black]
table {%
2.925 0.4631507775524
3.075 0.4631507775524
};
\addplot [black, mark=o, mark size=3, mark options={solid,fill opacity=0}, only marks]
table {%
3 0.311454172514304
3 0.504385964912281
};
\addplot [darkorange25512714]
table {%
0.85 0.465403913537213
1.15 0.465403913537213
};
\addplot [darkorange25512714]
table {%
1.85 0.46446895625322
2.15 0.46446895625322
};
\addplot [darkorange25512714]
table {%
2.85 0.41324987848584
3.15 0.41324987848584
};

\nextgroupplot[
height=\figureheight,
width=\figurewidth,
tick align=outside,
tick pos=left,
title={$WD$},
x grid style={darkgray176},
xmin=0.5, xmax=3.5,
xtick={1, 2, 3},
xticklabels={Baseline, TextTiling, TextSplit},
y grid style={darkgray176},
ymajorgrids,
ymin=0.286096682078403, ymax=0.689172757018301,
ytick style={color=black}
]
\path [draw=black, fill=pink]
(axis cs:0.85,0.480330194159717)
--(axis cs:1.15,0.480330194159717)
--(axis cs:1.15,0.596647439550724)
--(axis cs:0.85,0.596647439550724)
--(axis cs:0.85,0.480330194159717)
--cycle;
\addplot [black]
table {%
1 0.480330194159717
1 0.408888288531676
};
\addplot [black]
table {%
1 0.596647439550724
1 0.670851117248305
};
\addplot [black]
table {%
0.925 0.408888288531676
1.075 0.408888288531676
};
\addplot [black]
table {%
0.925 0.670851117248305
1.075 0.670851117248305
};
\path [draw=black, fill=lightblue]
(axis cs:1.85,0.438462518003814)
--(axis cs:2.15,0.438462518003814)
--(axis cs:2.15,0.485703468082345)
--(axis cs:1.85,0.485703468082345)
--(axis cs:1.85,0.438462518003814)
--cycle;
\addplot [black]
table {%
2 0.438462518003814
2 0.406093906093906
};
\addplot [black]
table {%
2 0.485703468082345
2 0.536064690026954
};
\addplot [black]
table {%
1.925 0.406093906093906
2.075 0.406093906093906
};
\addplot [black]
table {%
1.925 0.536064690026954
2.075 0.536064690026954
};
\addplot [black, mark=o, mark size=3, mark options={solid,fill opacity=0}, only marks]
table {%
2 0.304418321848399
};
\path [draw=black, fill=lightgreen]
(axis cs:2.85,0.395316507996345)
--(axis cs:3.15,0.395316507996345)
--(axis cs:3.15,0.429922973634411)
--(axis cs:2.85,0.429922973634411)
--(axis cs:2.85,0.395316507996345)
--cycle;
\addplot [black]
table {%
3 0.395316507996345
3 0.362402303578774
};
\addplot [black]
table {%
3 0.429922973634411
3 0.4631507775524
};
\addplot [black]
table {%
2.925 0.362402303578774
3.075 0.362402303578774
};
\addplot [black]
table {%
2.925 0.4631507775524
3.075 0.4631507775524
};
\addplot [black, mark=o, mark size=3, mark options={solid,fill opacity=0}, only marks]
table {%
3 0.311454172514304
3 0.504385964912281
};
\addplot [darkorange25512714]
table {%
0.85 0.504580362870291
1.15 0.504580362870291
};
\addplot [darkorange25512714]
table {%
1.85 0.481937176183607
2.15 0.481937176183607
};
\addplot [darkorange25512714]
table {%
2.85 0.41324987848584
3.15 0.41324987848584
};
\end{groupplot}

\end{tikzpicture}
		\caption[The $P_k$ and $WD$ scores achieved by the various segmenters]{The $P_k$ and $WD$ scores achieved by the various segmenters, compared with the baseline. The shown results are average values after 10 iterations.}
		\label{fig:results-box_plot}
\end{figure}
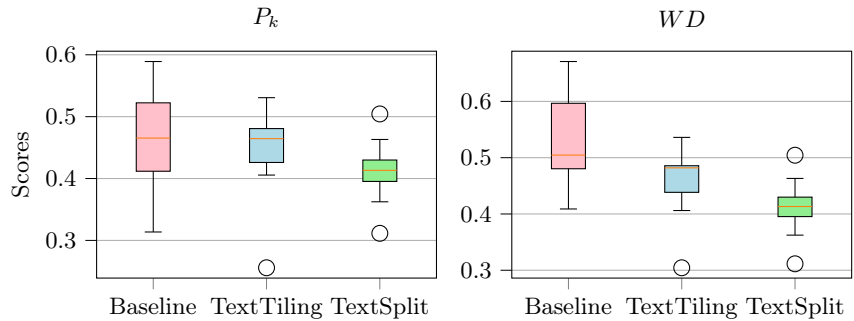

Based on the smallest produced $P_k$ and $WD$, we tune the parameters of the TextTiling and TextSplit algorithms using grid and linear search respectively. The optimal parameters were found to be $w=30$, $k=5$, and $f=0$ for the TextTiling algorithm and $l=10$ for the TextSplit algorithm. The effect of the utilised algorithms is illustrated in Figure \ref{fig:results-box_plot}, depicting their improvement over the baseline in terms of the $P_k$ and $WD$ error metrics. As it can be observed from the aforementioned figure, we conclude that the TextSplit algorithm (with $l=10$) achieved the lowest mean for both error metrics: $\bar{P_k}=0.41$ and $\bar{WD}=0.41$. We therefore employed its use for the second objective.

 \vspace{-1em}

\subsection{Text Summarisation}

\subsubsection{Adopted Evaluation}

Due to the absence of ground truth data, evaluation was carried out through the use of a preliminary survey \cite{rezapour2020spotify, karlbom2020abstractive}, whereby participants were questioned on the summary's relevancy with respect to its corresponding segment. The survey comprised of a small set of 10 randomly sampled podcast segments (one from each episode) extracted by the best segmentation algorithm (TextSplit, in this case). The three summaries generated by T5, BART, and Pegasus, alongside a human-written title\footnote{The human-written titles underwent scrutiny to ensure they accurately represented the content of the respective segment in a comprehensive manner.}, were included for each segment. A human-written title was also included. After listening to the audio segment, the respondents were requested to score each title (on a 5-Point Liktert Scale) based on the relevancy of the summary. Given $N$ participants, we define the average relevancy score $R_x$ as outlined in Equation 4.
\begin{equation}
    R_x = \frac{\sum^{\tilde{\mathcal{S}_x}}_{i=1}\textrm{score}(S_{i})}{|\tilde{\mathcal{S}_x|} \cdot N}
    \label{eqn-avg-relevancy}
\end{equation}
This allows the quantitative comparison of the generated summaries' quality and is calculated with respect to some text summarisation model $x$. The total score obtained on some summary $S_{i}$ is given by $\textrm{score}(S_i)$. The set of the sampled summaries generated by model $x$ is denoted by $\tilde{\mathcal{S}_x}$. A model that achieves a relevancy score equal to or higher than that of human-written titles is considered to be a favorable outcome.

\subsubsection{Results}

For each of the 10 sampled segments, we generate a summary using the considered text summarisation models and surveyed 25 human participants to assess their relevancy. The results from the survey revealed that the T5 model produced the best summaries ($R_\textrm{T5} = 3.34$), achieving a relevancy score closest to that of the human-written summaries ($R_\textrm{Human} = 3.64$). On the contrary, BART, and Pegasus ($R_\textrm{BART} = 2.60$ and $R_\textrm{Pegasus} = 2.30$ respectively) performed the worst due to the scores of the irrelevant summaries they sometimes produced. We analyse the effect of variables towards the relevancy scores in Table \ref{tab:corr}. For example, we investigate the initial intuition that an episode with high segmentation error consists of segments with poor topic coverage, and therefore, highly irrelevant summaries. From such an analysis, we note that no significant correlation between the quality of the episodes' segmentation and the summaries' relevancy was found. While the addition of more samples may be necessary to substantiate this finding, we conjecture that a segment's summary can still be considered relevant even if the segment does not adhere to a segmentation baseline. Additionally, segment length was positively correlated with higher relevancy for Pegasus, but not for the other models. While investigating this may fall outside the scope of this study, it is worth noting that the Pegasus model underwent fine-tuning using the AESLC corpus, which, in comparison to the datasets employed for fine-tuning BART and T5, exhibits the smallest average document size \cite{zhang2019email, calizzano2022generating}. Interestingly, longer titles generated by T5 was also attributed to a higher relevancy score for the BART model, and shorter titles generated by BART attributed to a higher relevancy score for the T5 model.

	\begin{table}[!htb]
		\begin{center}
			\begin{tabular}{| c | c c c c | }
                \hline
                & \multicolumn{4}{c|}{Relevancy} \\
				& Human & T5 & BART & Pegasus \\
				\hline
				Segment Length & .304 & -.032 & .242 & \cellcolor{mygreen} .708*\\
				Human Summary Length & \cellcolor{mygreen} .745* & .049 & -.07& .016\\
				T5 Summary Length & -.072 & -.388 & \cellcolor{mygreen}  .738* & .355\\
				BART Summary Length & .27 & \cellcolor{mygreen}  -.668* & .02 & -.135\\
				Pegasus Summary Length & -.224 & .466 & -.251 & -0.08 \\
				Episode $P_k$ & -.153 & .205 & .360 & .026\\
				Episode $WD$ & -.1 & .263 & .288 & .102 \\
				\hline
				
			\end{tabular}
		\end{center}
		
		\caption{Pearson's correlation coefficients between the relevancy scores and the outlined variables (* = significant at the p $<$ .05 level). }
		\label{tab:corr}
		
	\end{table}

\section{Conclusion}

In this research, we have explored the combined application of topic segmentation and text summarisation methods to investigate how podcast episode comprehension can be improved. Using the segments generated by the best performing topic segmenter (TextSplit with $\bar{P_k}=0.41$ and $\bar{WD}=0.41$), we employed the considered text summarisation models (T5, BART, and Pegasus) to produce respective summaries. A survey was rolled out to 25 human participants to assess the relevancy of the generated summaries. From the reported results, we deem the T5 model as the most promising text summarisation model out of the three. Having an average relevancy score of $3.34$, the T5 model was off by $8\%$ when compared to the human-written titles. In conclusion, by investigating the effectiveness and efficiency of topic segmentation alongside text summarisation techniques, we show that such combination can indeed improve podcast episode comprehension. A limitation of our approach is the small samples taken to evaluate the second objective. Future work may expand the total number of surveyed users and sampled episodes. For example, sampling multiple segments within an episode may provide further insights. Additionally, other algorithms that were not explored in our experiments can also be considered. The code used in this study is freely available on GitHub \cite{Diacono_Spotify_Text_Segmentation_2022}.

\printbibliography

\newpage

\begin{subappendices}
\renewcommand{\thesection}{\Alph{section}}%

\section{Generated Summaries}
	\label{app:gen_sum}
	
	\begin{longtable}{| c | p{0.18\textwidth} | p{0.18\textwidth} | p{0.18\textwidth} | p{0.18\textwidth} |}
				\hline
				 \textbf{Segments} & \multicolumn{4}{|c|}{\textbf{Titles}} \\ 
				\hline
				& Human & T5 & BART & Pegasus \\
				\hline
				Segment 1 & Eat something before an argument & Getting Into An Argument - Eat Something & Had to work really well for us is if we start to get heated walk away & The first question one of us would ask is when was the last time you ate \\
				\hline
				Segment 2 & Farmer eating dirty milk & A Cow Drop Some Milk Into A Field & I think I know what happened to a cow's milk & Milk \\
				\hline
				Segment 3 & Igor's Theme & I Got My Eyes Open - I Got My Eyes Open & Igor is Tyler, the Creator's first album in four years, is a story about the Alps and the people who live there & Album Review \\
			    \hline
			    \hline
			    Segment 4 & Target Audience of Human Library & Human Library - Is There a Target Audience? & Criterion Library, an open-air library in the Indian city of Bangalore, has been talking to the BBC's Geeta Pandey about its work and its target audience. & Human Library \\
			    \hline
			    Segment 5 & Hating soda and candy & What Are Some Things You Loved As a Kid and Now Not So Much Or Hate? & What are some things you loved as a kid that you now hate? & Soda\\
			    \hline

		\caption{Example list of generated summaries.}
		\label{app:tab:gen_sum}
		
        \end{longtable}

\end{subappendices}

\end{document}